\documentclass[traditabstract]{aa} 
\usepackage{savesym}
\usepackage{amsmath}
\savesymbol{iint}
\usepackage{graphicx}
\usepackage[varg]{txfonts}
\usepackage{natbib,twoopt}
\usepackage[breaklinks=f]{hyperref}
\usepackage{epsfig}
\usepackage{txfonts}
\usepackage{amssymb}
\usepackage{hyperref}
\usepackage{color}
\restoresymbol{TXF}{iint}
\bibpunct{(}{)}{;}{a}{}{,} 
 
\newcommandtwoopt{\citeads}[3][][]{\href{http://adsabs.harvard.edu/abs/#3}%
{\citealp[#1][#2]{#3}}}
\newcommandtwoopt{\citepads}[3][][]{\href{http://adsabs.harvard.edu/abs/#3}%
{\citep[#1][#2]{#3}}}
\newcommandtwoopt{\citetads}[3][][]{\href{http://adsabs.harvard.edu/abs/#3}%
{\citet[#1][#2]{#3}}} 
\newcommandtwoopt{\citeyearads}[3][][]%
{\href{http://adsabs.harvard.edu/abs/#3}{\citeyear[#1][#2]{#3}}}

\newcommand{\ssc}{\mathrm{SC}}
\newcommand{\slc}{\mathrm{LC}}

\newcommand{\pv}{\vec{\theta}}

\def\teff{$T_\mathrm{eff}$}                 
\def\ms{\hbox{\,m\,s$^{-1}$}}               
\def\m2s2{\hbox{\,m$^{2}$\,s$^{-2}$}}       
\def\kms{\hbox{\,km\,s$^{-1}$}}             
\def\gcm3{\hbox{\,g\,cm$^{-3}$}}            
\def\vsini{\hbox{$v$\,sin\,$i_{\star}$}}    
\def\Msun{\hbox{$M_{\odot}$}}               
\def\Rsun{\hbox{$R_{\odot}$}}               
\def\Mjup{\hbox{$\mathrm{M}_\mathrm{Jup}$}} 
\def\Rjup{\hbox{$\mathrm{R}_\mathrm{Jup}$}} 
\def\degr{\hbox{$^\circ$}}                  
\def\mp{{\emph M}$_\mathrm{p}$}             
\def\rp{{\emph R}$_\mathrm{p}$}             
\def\corot{\emph{CoRoT}}                    
\def\kepler{\emph{Kepler}}                  

\begin{document}

\title{Kepler-77b: a very low albedo, Saturn-mass transiting planet \\ around a metal-rich solar-like star\,\thanks{Based on observations obtained with the 2.1-m Otto Struve telescope at McDonald Observatory, Texas, USA.}$^,$\thanks{Based on observations obtained with the Nordic Optical Telescope, operated on the island of La Palma jointly by Denmark, Finland, Iceland, Norway, and Sweden, in the Spanish Observatorio del Roque de los Muchachos of the Instituto de Astrofisica de Canarias, in time allocated by OPTICON and the Spanish Time Allocation Committee (CAT).}$^,$\thanks{The research leading to these results has received funding from the European Community's Seventh Framework Programme (FP7/2007-2013) under grant agreement number RG226604 (OPTICON).}}


   \author{D.~Gandolfi\inst{\ref{ESA}}
      \and H.~Parviainen\inst{\ref{IAC},\ref{LaLaguna}}
      \and M.~Fridlund\inst{\ref{ESA}}
      \and A.\,P.~Hatzes\inst{\ref{Tautenburg}}
      \and H.\,J.~Deeg\inst{\ref{IAC},\ref{LaLaguna}}
      \and A.~Frasca\inst{\ref{OACT}}
      \and A.\,F.~Lanza\inst{\ref{OACT}}
      \and P.\,G.~Prada~Moroni\inst{\ref{UniPisa},\ref{INAFPisa}}
      \and E.~Tognelli\inst{\ref{UniPisa}}
      \and A.~McQuillan\inst{\ref{TelAviv}}
      \and S.~Aigrain\inst{\ref{Oxford}}
      \and R.~Alonso\inst{\ref{IAC},\ref{LaLaguna}}
      \and V.~Antoci\inst{\ref{SAC}}
      \and J.~Cabrera\inst{\ref{DLR}}
      \and L.~Carone\inst{\ref{KU-Leuven}}
      \and Sz.~Csizmadia\inst{\ref{DLR}}
      \and A.\,A.~Djupvik\inst{\ref{NOT}}
      \and E.\,W.~Guenther\inst{\ref{Tautenburg}}
      \and J.~Jessen-Hansen\inst{\ref{SAC},\ref{NOT}}
      \and A.~Ofir\inst{\ref{Goettingen}}
      \and J.~Telting\inst{\ref{NOT}}
          }

\institute{Research and Scientific Support Department, European Space Agency (ESA/ESTEC), PO Box 299, 2200 AG Noordwijk, The Netherlands\label{ESA}\\ 
           \email{dgandolf@rssd.esa.int}
      \and Instituto de Astrof\'\i sica de Canarias, C/\,V\'\i a L\'actea s/n, 38205 La Laguna, Spain\label{IAC}
      \and Departamento de Astrof\'\i sica, Universidad de La Laguna, 38206 La Laguna, Spain\label{LaLaguna}
      \and Th\"uringer Landessternwarte, Sternwarte 5, Tautenburg, D-07778 Tautenburg, Germany\label{Tautenburg}
      \and INAF - Osservatorio Astrofisico di Catania, Via S. Sofia, 78, 95123 Catania, Italy\label{OACT}
      \and Physics Department ``E. Fermi'' University of Pisa, largo B. Pontecorvo 3, 56127 Pisa, Italy\label{UniPisa}
      \and Istituto Nazionale di Fisica Nucleare, largo B. Pontecorvo 3, 56127 Pisa, Italy\label{INAFPisa}
      \and School of Physics and Astronomy, Raymond and Beverly Sackler Faculty of Exact Sciences, Tel Aviv Uni., Tel Aviv 69978, Israel\label{TelAviv}
      \and Department of Physics, University of Oxford, Oxford, OX1 3RH, United Kingdom\label{Oxford}
      \and Stellar Astrophysics Centre, Department of Physics and Astronomy, $\AA$rhus Uni., Ny Munkegade 120, DK-8000 $\AA$rhus C, Denmark\label{SAC}
      \and Institute of Planetary Research, German Aerospace Center, Rutherfordstrasse 2, 12489 Berlin, Germany\label{DLR}
      \and KU Leuven, Centre for mathematical Plasma Astrophysics, Celestijnenlaan 200B, 3001 Leuven, Belgium\label{KU-Leuven}
      \and Nordic Optical Telescope, Apartado 474, 38700, Santa Cruz de La Palma, Spain\label{NOT}
      \and Institut f\"ur Astrophysik, Georg-August-Universit\"at, Friedrich-Hund-Platz 1, 37077 G\"ottingen, Germany\label{Goettingen}
           }

\date{Received 16 May 2013 / Accepted 28 June 2013}

 
\abstract{We report the discovery of \object{Kepler-77b} (alias \object{KOI-127.01}), a Saturn-mass transiting planet in a 3.6-day orbit around a metal-rich solar-like star. We combined the publicly available \kepler\ photometry (quarters 1--13) with high-resolution spectroscopy from the Sandiford@McDonald and FIES@NOT spectrographs. We derived the system parameters via a simultaneous joint fit to the photometric and radial velocity measurements. Our analysis is based on the Bayesian approach and is carried out by sampling the parameter posterior distributions using a Markov chain Monte Carlo simulation. Kepler-77b is a moderately inflated planet with a mass of \mp~$=0.430\pm0.032$~\Mjup, a radius of \rp~$=0.960\pm0.016$~\Rjup, and a bulk density of $\rho_\mathrm{p}=0.603\pm0.055$~\gcm3. It orbits a slowly rotating ($P_\mathrm{rot}=36\pm6$~days) G5\,V star with $M_\star=0.95\pm0.04$~\Msun, $R_\star=0.99\pm0.02$~\Rsun, \teff~$=5520\pm60$~K, [M/H]\,$=0.20\pm0.05$\,dex, that has an age of $7.5\pm2.0$~Gyr. The lack of detectable planetary occultation with a depth higher than $\sim$10\,ppm implies a planet geometric and Bond albedo of $A_\mathrm{g}\le0.087\pm0.008$ and $A_\mathrm{B}\le0.058\pm0.006$, respectively, placing Kepler-77b among the gas-giant planets with the lowest albedo known so far. We found neither additional planetary transit signals nor transit-timing variations at a level of $\sim$0.5 minutes, in accordance with the trend that close-in gas giant planets seem to belong to single-planet systems. The 106 transits observed in short-cadence mode by \kepler\ for nearly 1.2 years show no detectable signatures of the planet's passage in front of starspots. We explored the implications of the absence of detectable spot-crossing events for the inclination of the stellar spin-axis, the sky-projected spin-orbit obliquity, and the latitude of magnetically active regions.}

               
\titlerunning{The transiting Saturn-mass planet Kepler-77b}
\authorrunning{Gandolfi et al.}

   \maketitle
%

\section{Introduction}
\label{Introduction}

Space-based transit surveys are opening up a new exciting~era in exoplanetary science, enlarging the known parameter space of planetary systems. The high-precision, long-term, and uninterrupted photometry from \corot\ \citep{Baglin2006} and \kepler\ \citep{Borucki2010} enables us to detect transiting planets down to the Earth- and Mercury-like regime (e.g., CoRoT-7b, \citeauthor{Leger2009}\,\citeyear{Leger2009}; Kepler-10b, \citeauthor{Batalha2011}\,\citeyear{Batalha2011}; Kepler-37b \citeauthor{Barclay2013}\,\citeyear{Barclay2013}), planets in long-period orbits (e.g., CoRoT-9b, \citeauthor{Deeg2010}\,\citeyear{Deeg2010}), planets in the habitable zone (e.g., Kepler-22b, \citeauthor{Borucki2012}\,\citeyear{Borucki2012}), multi-transiting systems (e.g., CoRoT-24, \citeauthor{Alonso2012}\,\citeyear{Alonso2012}; Kepler-11, \citeauthor{Lissauer2011}\,\citeyear{Lissauer2011}), and circumbinary systems (e.g., Kepler-16, \citeauthor{Doyle2011}\,\citeyear{Doyle2011}; Kepler-47, \citeauthor{Orosz2012}\,\citeyear{Orosz2012}). The exquisite photometry from space, combined with ground-based spectroscopy, enables us to derive the most precise planetary and stellar parameters to date, which in turn are essential for investigating the internal structure, formation, and migration of planets.

Based upon the analysis of the first sixteen months of \kepler\ photometry, \citet{Batalha2013} have recently announced 2321 transiting planet candidates (\kepler\ Object of Interests, KOIs), almost doubling the number of candidates previously published by \citet{Borucki2011}. Although those objects have been accurately vetted for astrophysical false-positives using \kepler\ photometry \citep[see, e.g., Sect.\,4 in][]{Batalha2013}, many of them - especially those where only a single planet is observed to transit - remain planetary candidates and deserve additional investigations, including high-resolution spectroscopy and radial velocity (RV) measurements. Different configurations involving eclipsing binary systems can indeed mimic a transit-like signal \citep{Brown2003}, and can only be weeded out with ground-based follow-up observations. Given the large number of KOIs, the \kepler\ team is currently focusing its spectroscopic follow-up effort on the smallest planets, while relying on the rest of the scientific community to confirm the remaining planetary objects. Ground-based observations \citep[e.g.,][]{Colon2012,Lillo-Box2012,Santerne2012} have recently proven that the false-positive rate of \kepler\ transiting candidates might be noticeably higher than the 5\,\% previously claimed by \citet{Morton2011b}, proving the need for more investigations to assess the real nature of these objects.

We herein announce the discovery of the \kepler\ transiting planet \object{Kepler-77b} (also known as \object{KOI-127.01}), a Saturn-mass gas giant planet orbiting a metal-rich solar-like star. We combined \kepler\ public photometry with high-resolution spectroscopy carried out with Sandiford@McDonald-2.1m and FIES@NOT to confirm the planetary nature of the transiting object and derive the system parameters.

The paper is structured as follows. In Sect.~\ref{Kepler-Phot}, we describe the \kepler\ data available at the time of writing. In Sect.~\ref{HighRes_Spec}, we report on our high-resolution spectroscopic follow-up of Kepler-77. In Sect.~\ref{Parent_Star}, we present the properties of the planet's host star. In Sect.~\ref{GlobalAnalysis}, we simultaneously model the photometric and spectroscopic data using a Bayesian-based model approach. Finally, in Sect.~\ref{Discussion}, we present the fundamental parameters of Kepler-77b, look for additional companions in the system, constrain the planet's albedo, and explore the limits placed by the apparent absence of spot-crossings on the inclination of the stellar spin-axis, the sky-projected spin-orbit obliquity, and the latitude of magnetically active regions.

\section{\kepler\ photometry}
\label{Kepler-Phot}

\begin{table}[!t]
\label{StarTable}      
  \caption{\kepler, KOI, GSC2.3, USNO-A2, and 2MASS identifiers of the planet-hosting star Kepler-77. 
           Equatorial coordinates and optical SDSS-$g$,-$r$,-$i$,-$z$ photometry are from
           the \kepler\ Input Catalogue. Infrared $J$,$H$,$Ks$ and $W1$,$W2$,$W3$,$W4$ data are
           taken from the 2MASS \citep{Cutri03} and WISE \citep{Wright2010} database, respectively.}
  \centering
  \begin{tabular}{lll}       
  \multicolumn{1}{l}{\emph{Main identifiers}}     \\
  \hline
  \hline
  \noalign{\smallskip}                
  Kepler~ID       & 8359498           \\
  KOI~ID          & 127               \\
  GSC2.3~ID       & N2K0000444        \\
  USNO-A2~ID      & 1275-11111739     \\
  2MASS~ID        & 19182590+4420435  \\
  \noalign{\smallskip}                
  \hline
  \noalign{\medskip}
  \noalign{\smallskip}                
  \multicolumn{2}{l}{\emph{Equatorial coordinates}}     \\
  \hline                                  
  \hline                                  
  \noalign{\smallskip}                
  RA \,(J2000)      & $19^h\,18^m\,25\fs910$         \\
  Dec (J2000)     & $+44\degr\,20\arcmin\,43\farcs51$  \\
  \noalign{\smallskip}                
  \hline
  \noalign{\medskip}
  \noalign{\smallskip}                
  \multicolumn{3}{l}{\emph{Magnitudes}} \\
  \hline
  \hline
  \noalign{\smallskip}                
  \centering
  Filter \,\,($\lambda_{\mathrm eff}$)& Mag         & Error  \\
  \noalign{\smallskip}                
  \hline
  \noalign{\smallskip}                
  $g$ \,\,~\,(~0.48\,$\mu m$) & 14.449       & 0.030 \\
  $r$ \,\,~\,(~0.63\,$\mu m$) & 13.871       & 0.030 \\
  $i$ \,\,~\,(~0.77\,$\mu m$) & 13.720       & 0.030 \\
  $z$ \,\,~\,(~0.91\,$\mu m$) & 13.658       & 0.030 \\
  $J$ \,\,~\,(~1.24\,$\mu m$) & 12.758       & 0.023 \\
  $H$ \,\,\,(~1.66\,$\mu m$)  & 12.444       & 0.019 \\
  $Ks$  \,(~2.16\,$\mu m$)    & 12.361       & 0.018 \\
  $W_1$ \,(~3.35\,$\mu m$)    & 12.309       & 0.023 \\
  $W_2$ \,(~4.60\,$\mu m$)    & 12.377       & 0.023 \\
  $W_3$ \,(11.56\,$\mu m$)    &$12.587^{a}$  &   ~~~-\\
  $W_4$ \,(22.09\,$\mu m$)    &$~~9.492^{a}$ &   ~~~-\\
  \noalign{\smallskip}                
  \hline
  \end{tabular}
   \tablefoot{\tablefoottext{a}{Upper limit}}
\end{table}

\object{Kepler-77}, whose main designations, equatorial coordinates, and optical and infrared photometry are reported in Table~\ref{StarTable}, was previously identified as a \kepler\ planet-hosting star candidate by \citet{Borucki2011} and \citet{Batalha2013} and listed as \object{KOI-127}. Speckle observations \citep{Howell2011} and adaptive optics imaging \citep{Adams2012} excluded a background eclipsing binary spatially close to the star.

At the time of writing, the public \kepler\ data for Kepler-77, i.e. quarters 1--13 (Q$_{1}$--Q$_{13}$)\footnote{Available at \url{archive.stsci.edu/kepler}.}, consist of more than three years of nearly continuous observations, from 13 May 2009 to 27 June 2012. Two hundred and ninety-eight planetary-like transits were observed with a long-cadence (LC) sampling time of $T_\mathrm{exp}$=1765.5 sec. One hundred and six transits were also observed with a short-cadence (SC) sampling time of $T_\mathrm{exp}$=58.89 sec, from quarter 3 (Q$_{3}$) to quarter 7 (Q$_{7}$). 

The photometric data were automatically processed with the new version of the \kepler\ pre-search data conditioning (PDC) pipeline \citep{Stumpe2012}, which uses a Bayesian maximum a posteriori (MAP) approach to remove the majority of instrumental artefacts and systematic trends \citep{Smith2012}. Clear outliers and photometric discontinuities across the data gaps that coincide with the quarterly rolls of the spacecraft were removed by visual inspection of the PDC-MAP light curve. We performed a test analysis of the aperture photometry produced by the \kepler\ pipeline, with and without using the PDC-MAP. We found that the data reduction performed by the automatic pipeline does not affect the planet characterisation for the current case and we used the PDC-MAP light curve for our analysis (Sect.~\ref{GlobalAnalysis}). The point-to-point scatter estimates for the PDC-MAP LC and SC data are 378 and 1398~ppm, respectively (Sect.~\ref{GlobalAnalysis}). 

\begin{figure*}[t] 
\begin{center}
\resizebox{\hsize}{!}{\includegraphics[angle=0]{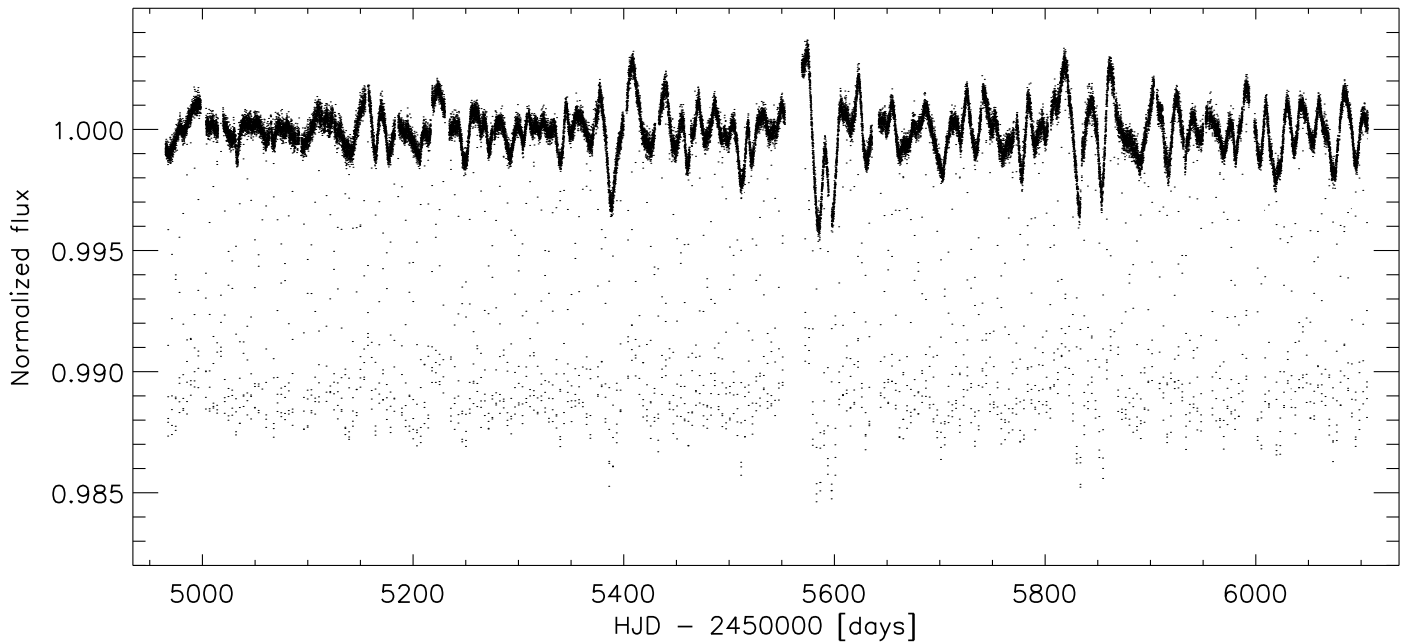}}
\caption{Q$_{1}$ -- Q$_{13}$ \kepler\ LC light curve of Kepler-77. The data from different quarters have been stitched together by arbitrarily normalising the flux by the out-of-transit median value per quarter. The points around normalised flux $\sim$0.99 mark the bottom of the 298 transits observed in LC mode. The light curve shows an out-of-transit flux variations with a peak-to-peak amplitude of $\sim$0.4\,\%.}
\label{LongCadence-LightCurve}
\end{center}
\end{figure*}

Figure~\ref{LongCadence-LightCurve} shows the cleaned PDC-MAP \kepler\ LC light curve of Kepler-77. The data from different quarters were stitched together by arbitrarily normalising the flux by the out-of-transit median value per quarter. The $\sim$1\%-deep transit signals are visible, along with a $\sim$0.4\,\% (peak-to-peak) out-of-transit variability whose nature is discussed in Sect.~\ref{Stellar_rotation}.

\section{High-resolution spectroscopy}
\label{HighRes_Spec}

\subsection{Sandiford observations}
\label{Sandidord-Obs}

We performed reconnaissance high-resolution spectroscopy with the Sandiford \'Echelle spectrometer \citep{McCarthy1993} attached at the Cassegrain focus of the 2.1-m telescope of the McDonald Observatory (Texas, USA). Four spectra were acquired at different epochs in May 2011, under clear and stable sky conditions. Three out of four observations were scheduled around orbital phases 0.25 and 0.75, i.e., when the highest amplitude of the RV curve is expected, to efficiently weed out an eclipsing binary scenario. We used the $1\arcsec$ slit width, which yields a resolving power of $R=47\,000$. As the size of the CCD can cover only $\sim$1000\,\AA, we set the spectrograph grating angle to encompass the wavelength range 5000--6000\,\AA. Taking into account the brightness of the target star, this was a good compromise between signal-to-noise (S/N) ratio and number of photospheric lines, which are suitable for RV measurements and reconnaissance spectral analysis. Three consecutive exposures of 1200\,sec were taken during each epoch observation to properly remove cosmic ray hits. We followed the observing strategy described in \citet{Buchhave2010}, i.e., we traced the RV drift of the instrument by acquiring long-exposed (T$_{\mathrm{exp}}$=30\,sec) ThAr spectra right before and after each epoch observation. The data were reduced using a customised IDL software suite, which includes bias subtraction, flat fielding, order tracing and extraction, and wavelength calibration. Radial velocity measurements were derived performing a multi-order cross-correlation with the RV standard star \object{HD\,182572} \citep{Udry1999}.

The Sandiford spectra revealed that Kepler-77 is a slowly rotating solar-like star. A first estimate of the photospheric parameters was made by using a modified version of the spectral analysis technique described in \citet{Gandolfi2008}, which is based on the use of a grid of stellar templates to simultaneously derive spectral type, luminosity class, and interstellar reddening from flux-calibrated low-resolution spectra. We modified the code to compare the co-added and normalised Sandiford spectrum with a grid of synthetic model spectra from \citet{Castelli2004}, \citet{Coelho2005}, and \citet{Gustafsson2008}. We found that Kepler-77 is a G5\,V star, with an effective temperature \teff\,$=5500\pm100$\,K, surface gravity log\,g\,$=4.50\pm0.10$~dex, metallicity [M/H]\,$=0.20\pm0.10$~dex, and a projected rotational velocity \vsini\,$\approx2$\,kms, in agreement with the photospheric parameters listed in the \kepler\ input catalogue\footnote{Available at: \\ \url{http://archive.stsci.edu/kepler/kepler_fov/search.php}.}.

The Sandiford RV measurements are listed in Table\,\ref{RV-Table}, along with the cross-correlation function (CCF) bisector spans, and the S/N ratio per pixel at 5500\,\AA. The reconnaissance Sandiford data reject a binary system scenario and are consistent with an RV variation of semi-amplitude $K\,\approx\,60$\,\ms\ in phase with the \kepler\ ephemeris (Fig.\,\ref{RV-Curve}). Assuming a mass of $M_\star\,\approx\,1$\,\Msun\ for the central star, as reported in the \kepler\ input catalogue, the RV semi-amplitude is compatible with a companion mass of \mp\,$\approx\,0.5$\,\Mjup. 

\begin{table}[t]
  \centering 
  \caption{Sandiford and FIES radial velocity measurements of Kepler-77. The CCF bisector spans and the S/N ratio per pixel at 5500\,\AA\ are listed in the last two columns.}
  \label{RV-Table}
\begin{tabular}{cccrc}
\hline
\hline
\noalign{\smallskip}
HJD              &   RV    & $\sigma_{\mathrm RV}$ &   Bisector  &  S/N/pixel    \\
($-$ 2\,450\,000)&   \kms  &    \kms               &     \kms    &  @5500\,\AA   \\
\noalign{\smallskip}
\hline
\noalign{\smallskip}
\multicolumn{1}{c}{Sandiford} \\
\noalign{\smallskip}
\hline
\noalign{\smallskip}
5701.82595 & $-$24.862  &         0.042         &  $-$0.032   &   16.2  \\
5703.80604 & $-$24.742  &         0.034         &     0.001   &   18.6  \\
5704.93367 & $-$24.818  &         0.042         &     0.011   &   15.4  \\
5705.78525 & $-$24.810  &         0.026         &  $-$0.004   &   22.2  \\
\noalign{\smallskip}
\hline
\noalign{\smallskip}
\multicolumn{1}{c}{FIES} \\
\noalign{\smallskip}
\hline
\noalign{\smallskip}
6101.69609 & $-$24.773  &         0.018         &     0.039   &  17.2  \\ 
6102.53160 & $-$24.808  &         0.018         &     0.002   &  20.8  \\
6104.60191 & $-$24.712  &         0.024         &     0.006   &  12.1  \\
6105.60914 & $-$24.805  &         0.020         &  $-$0.003   &  11.7  \\
6107.70122 & $-$24.691  &         0.019         &  $-$0.018   &  15.8  \\
6118.65614 & $-$24.705  &         0.014         &     0.028   &  20.5  \\
6119.70464 & $-$24.803  &         0.012         &  $-$0.024   &  24.2  \\
6121.66698 & $-$24.716  &         0.014         &     0.006   &  25.0  \\
6122.70204 & $-$24.702  &         0.023         &     0.067   &  11.1  \\
6202.41791 & $-$24.806  &         0.012         &     0.019   &  28.6  \\
6214.34702 & $-$24.739  &         0.024         &  $-$0.022   &  11.2  \\
\noalign{\smallskip}
\hline
\end{tabular}
\end{table}

\subsection{FIES observations}

The promising results obtained with Sandiford prompted us to continue the spectroscopic follow-up of Kepler-77. Eleven additional high-resolution spectra were obtained at different epochs, between June and October 2012, using the FIbre-fed \'Echelle Spectrograph \citep[FIES;][]{Frandsen1999} mounted at the 2.56-m Nordic Optical Telescope (NOT) of Roque de los Muchachos Observatory (La Palma, Spain). The observations were carried out under clear sky conditions, with seeing typically varying between 0.6 and 1.5\,\arcsec. We used the \emph{high-res} $1.3\,\arcsec$ fibre, which provides a resolving power of R=67\,000 and a wavelength coverage of about 3600--7400\,\AA. We adopted the same calibration scheme and exposure time as for the Sandiford follow-up. Observations were first scheduled around quadrature, to validate the trend inferred with Sandiford. Once the RV variation was confirmed, FIES RVs measurements were secured at different orbital phases to evenly cover the RV curve. The data were reduced in the same fashion as the Sandiford spectra. Once more, HD\,182572 was used as the RV standard star.  

\begin{figure}[t] 
 \begin{center}
\resizebox{\hsize}{!}{\includegraphics[angle=0]{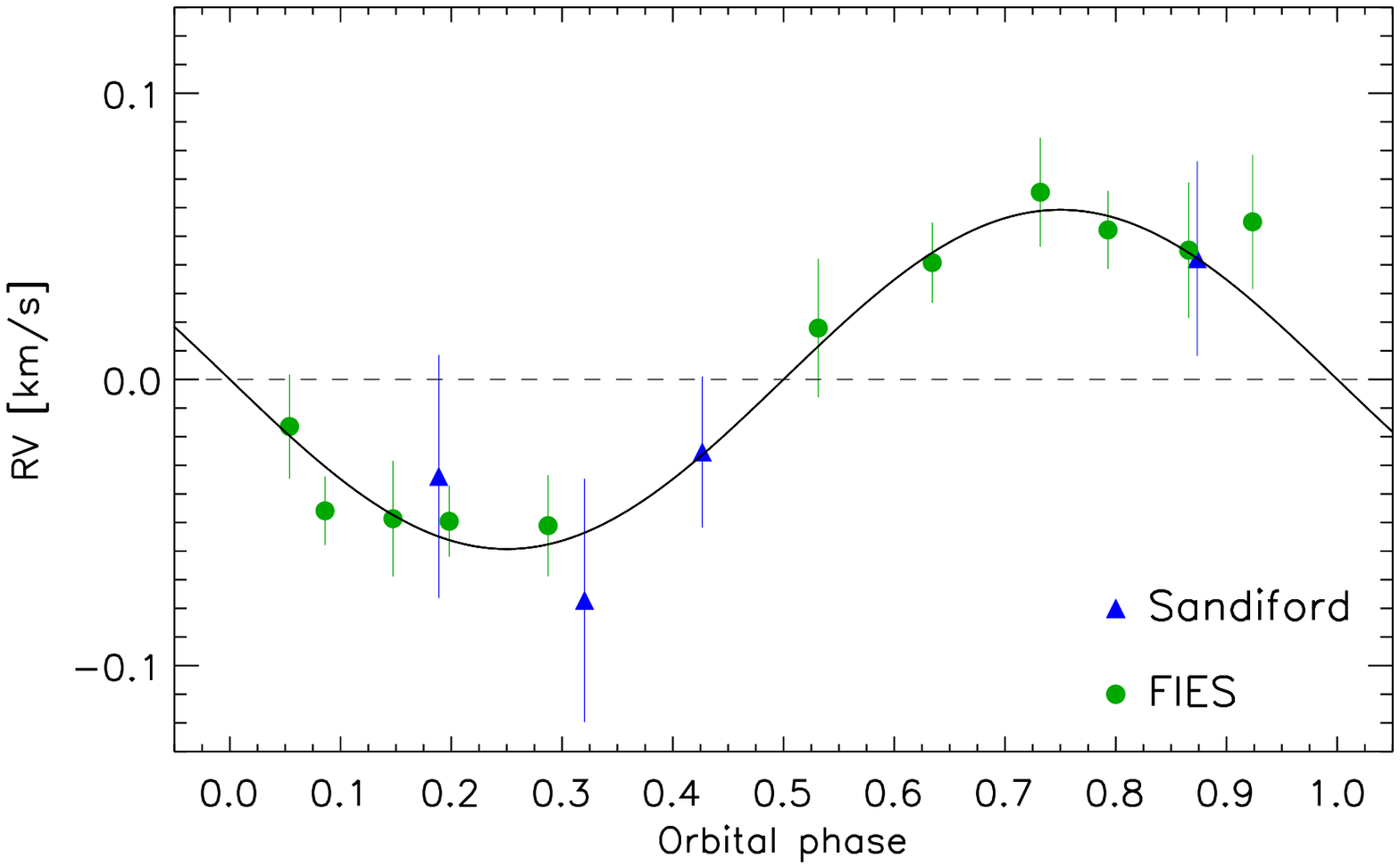}}
\resizebox{\hsize}{!}{\includegraphics[angle=0]{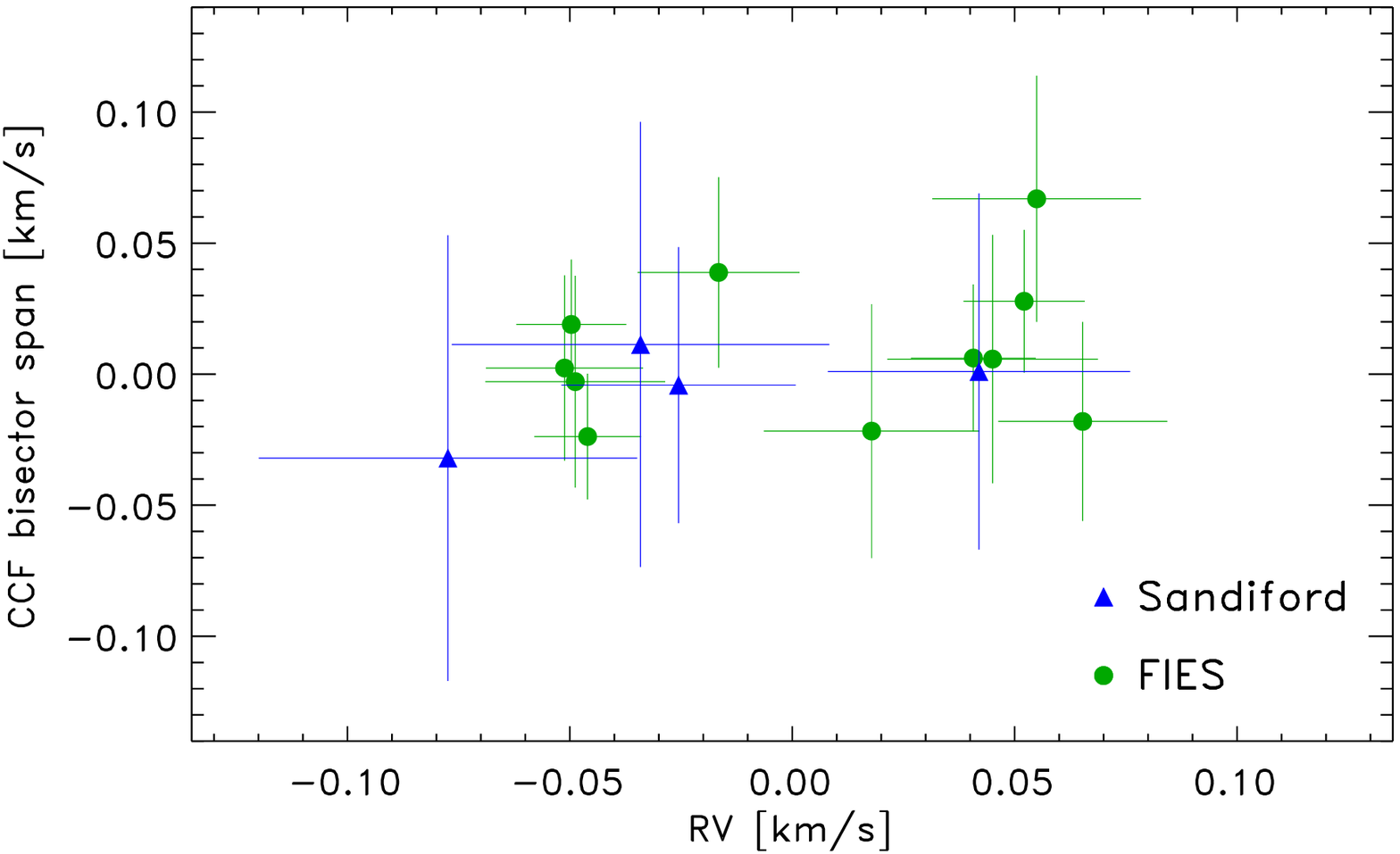}}
\caption{\emph{Upper panel:} Sandiford and FIES radial velocities measurements of Kepler-77 and Keplerian fit to the data. The systemic velocities for each instrument, as derived from the global modelling of the data (Sect.~\ref{GlobalAnalysis} and Table~\ref{Par_table}), have been subtracted from the RV measurements. \emph{Lower panel:} Bisector spans of the Sandiford and FIES cross-correlation functions versus RV measurements, after subtracting the systemic velocities. The error bars in the CCF bisector spans are taken to be twice the uncertainties in the RV data.}
\label{RV-Curve}
\end{center}
\end{figure}

The FIES RV measurements, CCF bisector spans, and S/N ratio per pixel at 5500\,\AA\ are listed in Table\,\ref{RV-Table}. The upper panel of Figure\,\ref{RV-Curve} shows the phase-folded RV curve of Kepler-77, as obtained from the global analysis of the photometric and spectroscopic data described in Sect.~\ref{GlobalAnalysis}.

Following the method outlined in \citet{Queloz01}, we performed an analysis of the Sandiford and FIES CCF bisector spans to exclude the presence of periodic distortions in the spectral line profile that might be caused either by stellar magnetic activity (i.e., photospheric spots and plages), or by a blended eclipsing binary whose light is diluted by Kepler-77. The lack of correlation between the CCF bisector span and RV data (Fig.\,\ref{RV-Curve}, lower panel) proves that the Doppler shift observed in Kepler-77 are induced by the companion's orbital motion.

\section{Properties of the parent star}
\label{Parent_Star}

\subsection{Photospheric parameters}
\label{Photosphericparameters}

An improvement on the determination of the photospheric parameters of Kepler-77 was achieved using the co-added FIES spectrum, which has a S/N ratio of about 65 per pixel at 5500\,\AA. We used the same procedure outlined in Sect.~\ref{Sandidord-Obs}, as well as the spectral analysis package SME\,2.1 \citep{Valenti1996,Valenti2005} and the ROTFIT code from \citet{Frasca2003,Frasca2006}. The latter is based on the use of ELODIE archive spectra of standard stars with well-known parameters \citep{Prugniel2001}. The three methods provided consistent results within the error bars. The microturbulent $v_ {\mathrm{micro}}$ and macroturbulent $v_{\mathrm{macro}}$ velocities were derived using the calibration equations of \citet{Bruntt2010} for Sun-like dwarf stars. The projected rotational velocity \vsini\ was measured by fitting the profile of several clean and unblended metal lines. The final adopted values are \teff\,$=5520\pm60$\,K, log\,g$=4.40\pm0.10$\,dex, [M/H]~$=0.20\pm0.05$\,dex, $v_ {\mathrm{micro}}=0.9\pm0.1$\,\kms, $v_{\mathrm{macro}}=1.8\pm0.3$\,\kms, and \vsini\,$=1.5\pm1.0$\,\kms, in very good agreement with the values derived from the Sandiford spectra and those reported in the \kepler\ input catalogue.

\subsection{Stellar mass, radius, and age}
\label{StellarMassRadiusAge}

We determined the mass $M_\star$ and radius $R_\star$ of the planet-hosting star Kepler-77 using the spectroscopically derived effective temperature \teff\ and metallicity [M/H], along with the bulk stellar density $\rho_\star$, as obtained from the transit light curve modeling (Sect.~\ref{GlobalAnalysis}). We compared the position of the star on a $\rho_\star$ versus \teff\ diagram with a grid of \emph{ad hoc} evolutionary tracks. 

Stellar models were generated using an updated version of the \emph{FRANEC} code \citep{DeglInnocenti2008,Tognelli2011} and adopting the same input physics and parameters as those used in the Pisa Stellar Evolution Data Base for low-mass stars\footnote{Available at: \\ \url{http://astro.df.unipi.it/stellar-models/}} \citep[see, e.g.,][for a detailed description]{DellOmodarme2012}. To account for the current photospheric metallicity of Kepler-77 ([M/H]\,$=0.20\pm0.05$\,dex) and microscopic diffusion of heavy elements towards the centre of the star (see below), we computed evolutionary tracks assuming an initial metal content of Z=0.0178, Z=0.0198, Z=0.0221, Z=0.0245, Z=0.0272, and Z=0.0301. The corresponding initial helium abundances (i.e., Y=0.284, 0.288, 0.293, 0.297, 0.303, and 0.308) were determined assuming a helium-to-metal enrichment ratio $\Delta$Y/$\Delta$Z=2 \citep{Jimenez2003,Casagrande2007,Gennaro2010} and a cosmological $^4$He abundance Y$_\mathrm{p}$=0.2485 \citep{Cyburt2004,Peimbert2007a,Peimbert2007b}. For each chemical composition, we generated a very fine grid of evolutionary tracks in the mass domain $M_\star=0.80$--$1.10$\,\Msun, with step of $\Delta M_\star=0.01$\,\Msun. Thus, a total of 186 stellar tracks were calculated specifically for this project.

As mentioned, the models take into account microscopic diffusion, where diffusion velocities for the different chemical species due to gravitational settling and thermal diffusion are calculated by means of the routine developed by \citet{Thoul1994}. Due to this effect, the surface chemical composition changes with time depending on the stellar mass. As a consequence, the metallicity currently observed on the surface of low-mass main-sequence stars is different from the original one, a difference that increases with the age. To improve the inferred stellar parameters resulting from the comparison between models and observational data, the effect of this change in the surface chemical composition has to be taken into account, as far as old low-mass main-sequence stars are analysed. 

To reproduce the current surface metallicity measured for Kepler-77, we found that evolutionary tracks with initial metal content between Z=0.0221 and Z=0.0272 had to be used. We derived a mass of $M_\star=0.95\pm0.04$\,\Msun\, a radius of $R_\star=0.99\pm0.02$\,\Rsun\, and an age of $7.5\pm2.0$\,Gyr (Table~\ref{Par_table}). Stellar mass and radius translate into a surface gravity of log\,g~$=4.42\pm0.01$~dex, which agrees very well with the spectroscopically derived value (log\,g~$=4.40\pm0.10$~dex). We note that assuming an initial metal content equal to the measured current value would lead to an underestimate of the stellar mass and radius of about 0.04\,\Msun\ and 0.01\,\Rsun\, respectively, as well as an overestimate of the age of about 2\,Gyr.

We also used Equation~32 from \citet{Barnes2010b} and the rotation period of the star ($P_\mathrm{rot}=36\pm6$~days; see Sect.~\ref{Stellar_rotation}) to infer the gyro-age of Kepler-77, assuming a convective turnover time-scale of $\tau_\mathrm{c} = 41.40$~days \citep{Barnes2010a} and a zero-age main-sequence rotation period of $P_0=1.1$\,days \citep{Barnes2010b}. We found a gyro-age of $7.3\pm2.4$\,Gyr, in very good agreement with the value obtained from the evolutionary tracks.

\subsection{Interstellar extinction and distance}
\label{Ext_Dist}

The interstellar extinction $A_\mathrm{v}$ and spectroscopic distance $d$ to Kepler-77 were obtained by applying the method described in \citep{Gandolfi2010}. Briefly, we simultaneously fitted the available SDSS, 2MASS, and WISE colours (Table~\ref{StarTable}) with synthetic theoretical magnitudes. These were obtained by integrating a \emph{NextGen} model spectrum \citep{Hauschildt99} - with the same photospheric parameters as the star - over the response curve of the SDSS, 2MASS, and WISE photometric systems. We excluded the $W_3$ and $W_4$ WISE magnitudes, which are only upper limits. Assuming a normal extinction ($R_\mathrm{v}=3.1$) and a black body emission at the stellar effective temperature and radius, we found that Kepler-77 suffers a moderately low interstellar extinction $A_\mathrm{v}=0.08\pm0.04$\,mag and that its distance is $d=570\pm70$\,pc (Table~\ref{Par_table}).

\subsection{Stellar rotation}
\label{Stellar_rotation}

The PDC-MAP LC light curve of Kepler-77 shows periodic and quasi-periodic flux variations with a peak-to-peak amplitude of $\sim$0.4\,\% (Fig.~\ref{LongCadence-LightCurve}). Given the G5\,V spectral of the star, these features are most likely caused by stellar activity, e.g., rotational modulation, differential rotation, and spot evolution. Unfortunately, we were unable to spectroscopically determine the magnetic activity level of Kepler-77, as the low S/N ratio of the blue end of the FIES spectra does not allow one to detect any significant emission in the core of the Ca\,{\sc ii} H and K lines.

\begin{figure}[t] 
\begin{center}
\resizebox{\hsize}{!}{\includegraphics[angle=0]{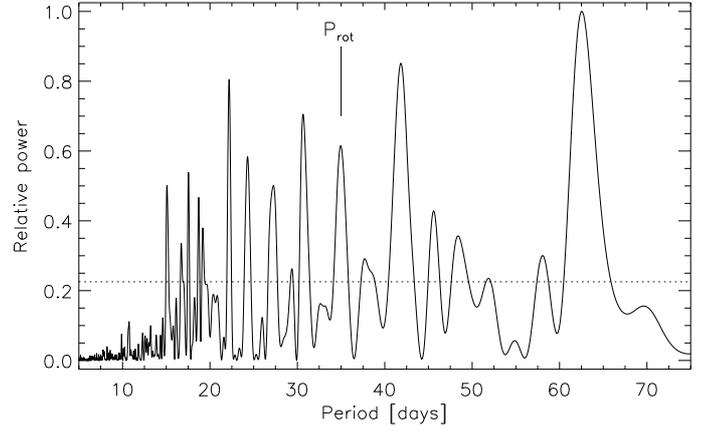}}
\caption{Lomb-Scargle periodogram of the \kepler\ LC light curve of Kepler-77, following subtraction of the best-fitting transit model. The horizontal dashed line denotes the 0.1\,\% false-alarm probability. The vertical line marks the rotational period of Kepler-77, as derived from the autocorrelation function method (see text for details).}
\label{Peridogram}
\end{center}
\end{figure}

The rotation period $P_\mathrm{rot}$ of a magnetically active star can be determined by searching for periodicities induced by starspots moving in and out of sight as the star rotates. Figure~\ref{Peridogram} shows the Lomb-Scargle periodogram of the \kepler\ LC light curve of Kepler-77, following subtraction of the best-fitting transit model (see Sect.~\ref{GlobalAnalysis}). There are at least nine significant periods in the range $15$--$65$~days with a false-alarm probability lower than 0.1\,\%. These are most likely due to spot evolution or harmonics of the rotational period, which makes it difficult to derive the true rotational period of Kepler-77.

Following the guidelines described in \citet{McQuillan2013}, we applied the autocorrelation function (ACF) method to derive the rotational period of Kepler-77. This technique measures the degree of self-similarity of the light curve over a range of lags. In the case of rotational modulation, a peak in the ACF will occur at the number of lags corresponding to the period at which the spot-crossing signature repeats. 

\begin{figure*}[th] 
\begin{center}
\resizebox{\hsize}{!}{\includegraphics[angle=0]{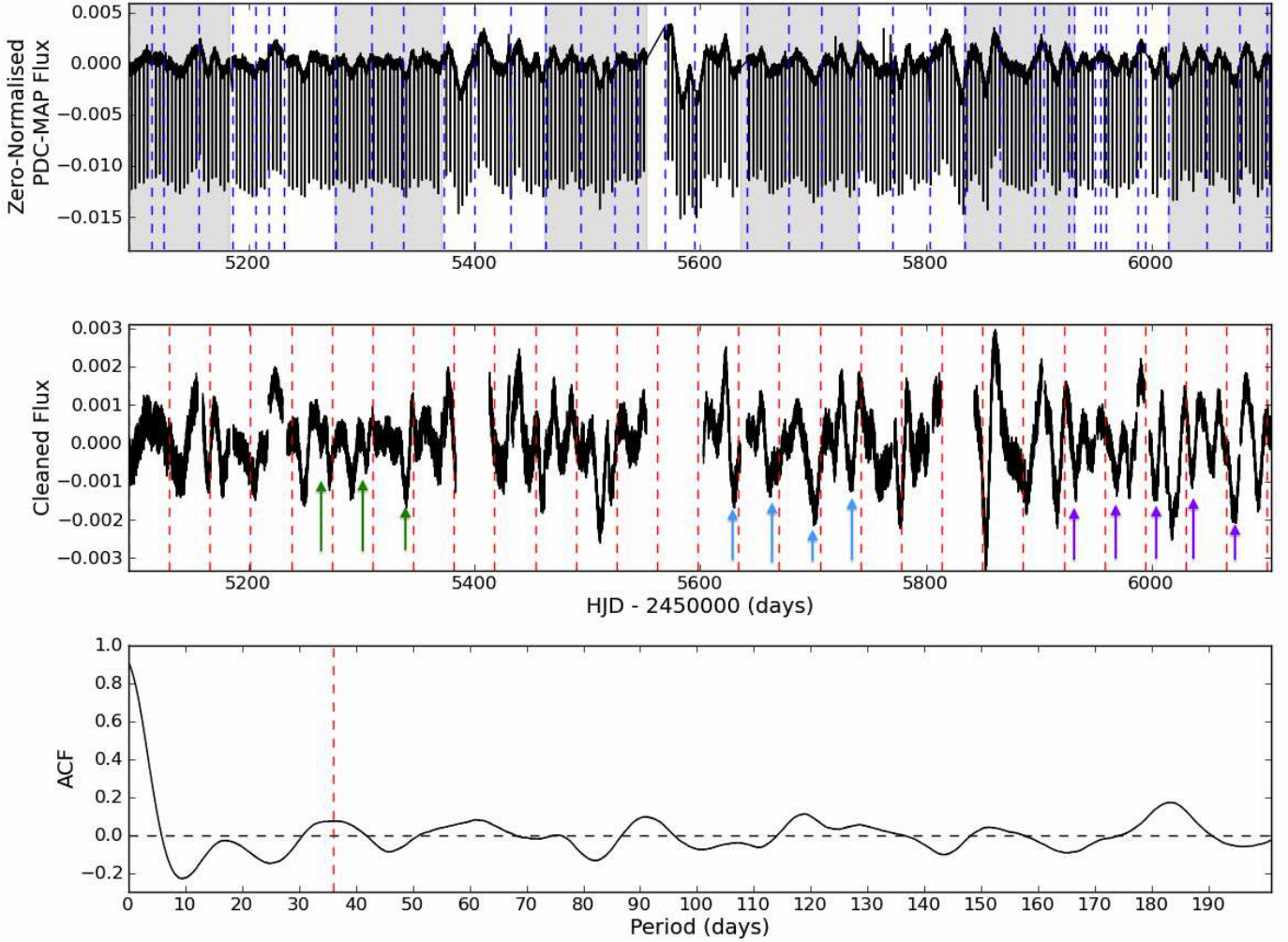}}
\caption{\emph{Upper panel}. PDC-MAP LC light curve of Kepler-77, with the location of known instrumental systematic features marked by blue dashed lines. Grey and white bands denote the zero-normalised Q$_{3}$--Q$_{13}$ data. \emph{Middle panel}. Cleaned and zero-normalised light curve, following subtraction of the best-fitting transit model. Rotational period intervals are shown with red dashed lines. The arrows mark groups (per colour) of repeated features. \emph{Lower panel}. Autocorrelation function (ACF) of the cleaned and zero-normalised light curve. The red dashed line marks the peak corresponding to the 36-day rotational period of Kepler-77.}
\label{Autocorrelation}
\end{center}
\end{figure*}

Although the PDC-MAP algorithm removes systematic instrumental fluctuations correlated across different stars in the same CCD channel, we cannot rule out residual instrumental effects and stochastic errors superimposed on the stellar variability (Figure~\ref{Autocorrelation}, upper panel). A visual inspection of the raw and PDC-MAP data resulted in the identification of three regions of the light curve with anomalously high amplitude variability that does not look astrophysical in shape. These regions occupy the in HJD - 2\,450\,0000 time intervals between 5383 -- 5413,  5563 -- 5603, and 5813 -- 5843 days (Figure~\ref{Autocorrelation}, middle panel). They were masked when calculating the ACF to minimise the effect of residual systematics, and, like other gaps in the light curve, they were set to zero to prevent them contributing to the ACF. Q$_{1}$ and Q$_{2}$ were also excluded due to the short length of Q$_{1}$ and the considerable contamination by instrumental faults in Q$_{2}$.

The ACF of the cleaned, best-fit transit model-subtracted light curve displays correlation peaks at $\sim$17.5 and $\sim$36 days (Figure~\ref{Autocorrelation}, lower panel). The latter is the dominant peak, and we attributed the peak at 17.5 days to a partial correlation between active regions at opposite longitudes of the star. To estimate the period and its error, we fitted a Gaussian function to the correlation peak and found $P_\mathrm{rot}=36\pm6$~days. The projected rotational velocity \vsini\,$=1.5\pm1.0$\,\kms\ and the stellar radius $R_\star=0.99\pm0.02$\,\Rsun\ give an upper limit on the rotational period of $33.4\pm11.1$~days, which agrees with our $P_\mathrm{rot}$ estimate.

Repeated features visible in the light curve, such as those marked with the arrow sets in the middle panel of Figure~\ref{Autocorrelation}, lead us to believe that the detected period results from spot modulation of the light curve.

\section{Photometric and spectroscopic global analysis}
\label{GlobalAnalysis}

We characterised the planet and its orbit via a simultaneous joint fit to the light curve and RV measurements using methods based on Bayesian statistics.

The \kepler\ photometric data included in the analysis is a subset of the whole light curve. We selected 12 hours of data-points around each transit and de-trended the individual transit light curves using a second-order polynomial fitted to the out-of-transit points. We preferred SC light curves when available, and excluded the LC transits for which SC data was available. The final SC and LC light curves contain $\sim$76800 and $\sim$4500 points, respectively.

We used a Bayesian approach for the parameter estimation, and sampled the parameter posterior distributions using Markov chain Monte Carlo (MCMC). The log-posterior probability was described as
\begin{equation}
\begin{split}
\log P(\vec{\theta}|D) = & \log P(\vec{\theta})  \\
&+ \log P(V_\mathrm{FIES}|\pv) + \log P(V_\mathrm{Sand}|\pv) \\
&+ \log P(F_\ssc|\pv) + \log P(F_\slc|\pv)\,, 
\end{split}
\label{Log-post-prob}
\end{equation}
where $V_\mathrm{FIES}$ and $V_\mathrm{Sand}$ correspond to the FIES and Sandiford radial velocity data, $F_\ssc$ and $F_\slc$ are the short- and long-cadence photometric data, $\pv$ is the parameter vector, and $D$ the combined dataset. The first term in the right-hand side of Equation~\ref{Log-post-prob}, namely $\log P(\vec{\theta})$, represents the logarithm of the joint prior probability (the product of individual parameter prior probabilities), and the four remaining terms are the likelihoods for the radial velocity and light curve data. The likelihoods of the combined dataset $D$ follow the basic form for a likelihood assuming independent identically distributed errors following normal distribution:
\begin{equation}
\begin{split}
 \log P(D|\pv) = & -\frac{N_\mathrm{D}}{2} \log(2\pi) - N_\mathrm{D} \log(\sigma_\mathrm{D}) \\
&- \sum_{i=1}^{N_\mathrm{D}} \left ( \frac{D_i - M(t_i,\vec{\theta})}{\sigma_\mathrm{D}} \right )^2\,,
\end{split}
\end{equation}
where $D_i$ is the single observed data point $i$, $M(t_i,\vec{\theta})$ the model explaining the data, $t_i$ the centre time for a data point $i$, $N_\mathrm{D}$ the number of data points and $\sigma_\mathrm{D}$ the standard deviation of the error distribution \citep[a more detailed derivation can be found, e.g., in][]{Gregory2005}.

The radial velocity model follows from equation 
\begin{equation}
 V = V_{\gamma} + K[\cos(\omega + \nu) + e\cos\omega]\,,
\end{equation}
where $V_{\gamma}$ is the systemic velocity, $K$ the radial velocity semi-amplitude, $\omega$ the argument of periastron, $\nu$ the true anomaly, and $e$ the eccentricity. To account for the RV offset between Sandiford and FIES, we used two separate systemic velocities for the two different datasets, but did not consider linear (or higher order) velocity trends.

\begin{figure}[t]
 \centering
 \resizebox{\hsize}{!}{\includegraphics[angle=0]{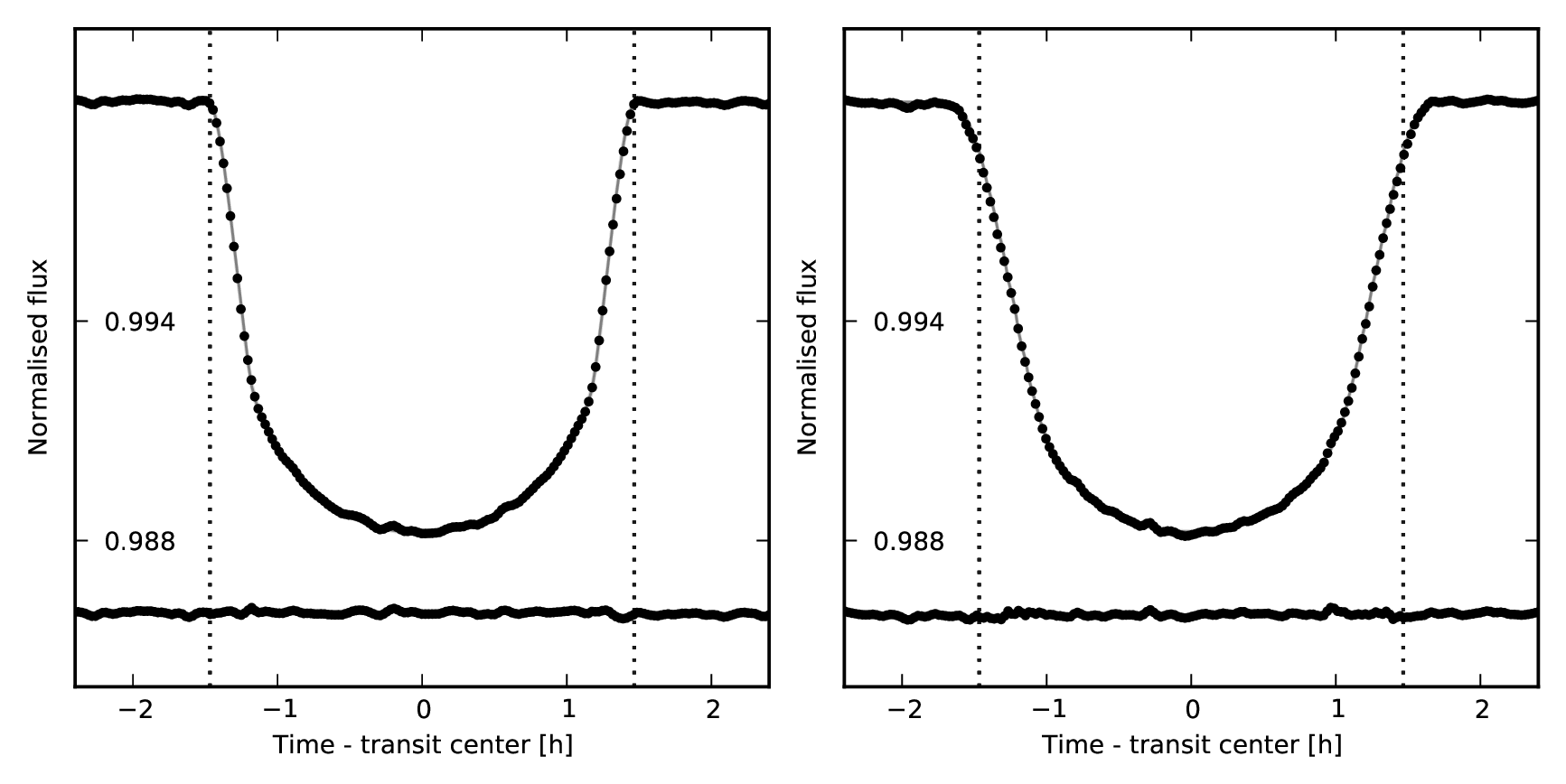}}
 \caption{Phase-folded transit light curve of Kepler-77. Short- and long-cadence \kepler\ data are shown in the left and right panel, respectively, binned at $\sim$1.7 minutes. The best-fitting model is overplotted with a solid grey line. The dashed vertical lines mark the planet's true first and fourth contact. The apparent transit duration is longer for the LC data than for the SC data, owing to the low time resolution.}
 \label{TransitLightcurve}
\end{figure}

The transit model used our implementation of the transit-shape model described in \citet{Gimenez2006} and optimised for light curves with a large number of data points\footnote{The code is freely available at \\ \url{https://github.com/hpparvi/PyTransit}.}. The long-cadence model was super-sampled using ten subsamples per LC exposure to reduce the effects from the extended integration time, a necessity noted by \citet{Kipping2010}. The widening of the ingress and egress durations is clearly visible in the phase-folded long-cadence transit light curve plotted in Fig.~\ref{TransitLightcurve} (right panel).

The final joint model had 12 free parameters: orbital period $P_\mathrm{orb}$, the epoch of mid-transit $T_\mathrm{c}$, transit duration $T_{14}$, planet-to-star area ratio $R_\mathrm{P}^{2}/R_{\star}^{2}$, impact parameter \hbox{$b=a_\mathrm{P}\,cos\,(i_\mathrm{P})/R_{\star}$}, linear $u_1$ and quadratic $u_2$ limb-darkening coefficients, SC photometric scatter $\sigma_\mathrm{SC}$, LC photometric scatter $\sigma_\mathrm{LC}$, radial velocity semi-amplitude $K$, and two systemic velocities, i.e, one for Sandiford $V_{\gamma\,\mathrm{Sand}}$ and one for FIES $V_{\gamma\,\mathrm{FIES}}$. When fitting for a non-zero orbital eccentricity, two additional parameters were included, namely, the eccentricity $e$ and argument of periastron~$\omega$.

We used uninformative priors (uniform) on all parameters. An initial fit for an eccentric orbit yielded $e=0.10^{+0.11}_{-0.07}$ and $\omega = 104.6^{+72.8}_{-31.4}$ degrees. We note that the derived non-zero eccentricity is lower than a $2\,\sigma$ detection. Furthermore, the system parameters derived for an eccentric orbit are well within $1\,\sigma$ from those obtained for a circular solution. Following the \emph{F-test} described in \citet{Lucy1971}, we found that there is a 66\,\% probability that the best-fitting eccentric solution could have arisen by chance if the orbit were actually circular. We thus decided to adopt a circular orbit ($e=0$). This assumption is also corroborated by the time-scale of orbital eccentricity dampening due to tidal interactions with the host star ($\tau_e=-e/\dot{e}$). Following the formalism of, e.g., \citet{Yoder1981} and \citet{Matsumura2010}, and assuming tidal dissipation quality factors $Q_\mathrm{p}\approx 10^4-10^6$ for gas giant planets \citep[e.g.,][]{Carone2012}, we found that $\tau_e=-e/\dot{e} \leq 0.5$~Gyr, i.e., at least one order of magnitude younger than the age of the system ($7.5\pm2.0$\,Gyr). Moreover, the negative search for other companions in the system makes it unlikely that a perturber is continually stirring up the eccentricity of Kepler-77b (Sect.\,\ref{AdditionalPlanets}). We can thus safely assume that a hypothetical initial eccentric orbit of Kepler-77b has most likely been circularised by tides in the past.

We carried out the sampling of the posterior distribution with \emph{emcee} \citep{Foreman-Mackey2012}, a Python implementation of the affine invariant Markov chain Monte Carlo sampler \citep{Goodman2010}, which offers excellent sampling properties. The sampling was carried out using 500 walkers (chains). We first ran the sampler iteratively through a burn-in period consisting of 10 runs of 300 steps each, after which the walkers had converged to the posterior distribution. The final sample consists of 500 iterations with a thinning factor of 10, leading to 25\,000 independent posterior samples. The affine invariant sampler is good at sampling correlated parameter spaces, and the fitting parameter set is designed to reduce the correlations (except when linear and quadratic limb-darkening coefficients are used directly instead of their linear combinations, as is usually done). We chose a thinning factor of 10 based on the parameter autocorrelation lengths to ensure that the samples can be considered to be independent. 

The best-fitting system parameters were taken to be the median values of the posterior probability distributions. Error bars were defined at the 68\,\% confidence limit. We summarise our results in Table~\ref{Par_table} and show the radial velocity and photometric data along with the best-fitting models in Figs.\ref{RV-Curve} and \ref{TransitLightcurve}, respectively.

\section{Discussion and conclusion}
\label{Discussion}

\subsection{Properties of the planet}

With a mass of \mp\,$=0.430\pm0.032$~\Mjup, radius of \rp\,$=0.960\pm0.016$~\Rjup, and bulk density of $\rho_\mathrm{p}=0.603\pm0.055$\,\gcm3, Kepler-77b joins the growing number of bloated gas-giant planets on a $\sim$3-day orbit. We used the empirical radius calibration for Saturn-mass planets found by \citet{Enoch2012} to estimate the predicted radius of Kepler-77b, given the planetary mass $M_\mathrm{p}$, equilibrium temperature $T_\mathrm{eq}$, semi-major axis $a_\mathrm{p}$, and stellar metallicity [M/H] (Table~\ref{Par_table}). Assuming a circular orbit - therefore omitting the contribution from tidal-dissipation - we found a predicted radius of $1.10\pm0.13$\,\Rjup, which agrees within $1\,\sigma$ with the measured values. Taking into account the relatively high metallicity of the star [M/H]\,$=0.20\pm0.05$\,dex, this further more proves that the radii of Saturn-mass planets ($\sim$0.1--0.5\,\Mjup) are almost exclusively dependent on their mass and heavy-element content \citep[see equation 7 in][]{Enoch2012}, with the second parameter affecting the mass of their cores \citep{Guillot2006}.

\begin{figure}[t]
\centering
\resizebox{\hsize}{!}{\includegraphics[angle=0]{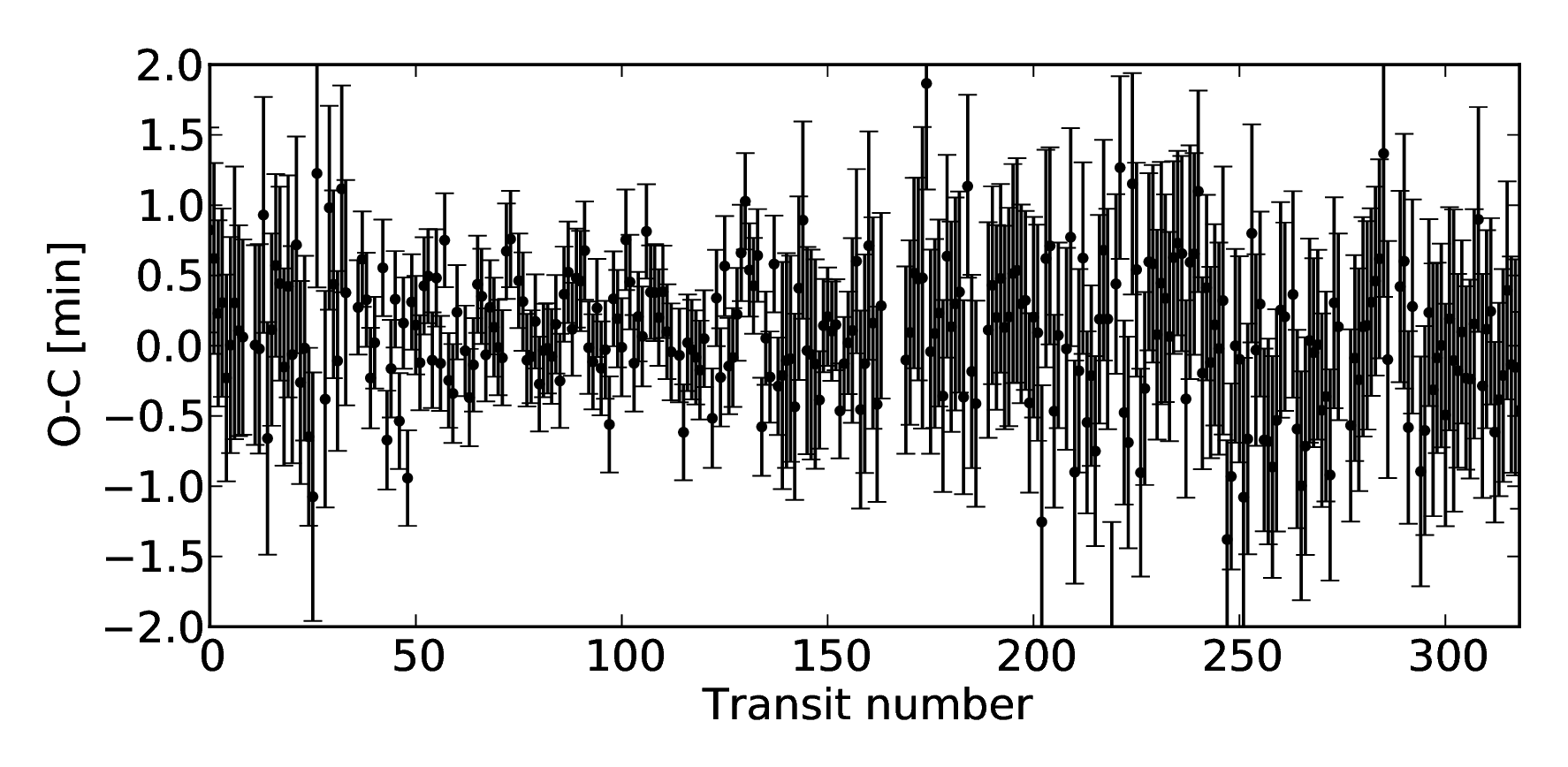}}
\caption{Differences between the observed and modelled transit centre times. The residuals show some structures, but they are only marginally significant.}
\label{KOI127_TTVs}
\end{figure}

\subsection{Search for planet occultation: the albedo of Kepler-77b}
\label{SecondaryTransit}

The superb photometric quality of \kepler\ light curves allows one to detect secondary eclipses with very high accuracy and measure the planet albedo. Following the method described in \citet{Parviainen2013}, we carried out a search for occultations of Kepler-77b by its host star. The search was performed at different orbital phases to allow for an eccentric orbit. Briefly, we modified the model used for the system characterisation to include the secondary eclipse - which was parameterised by the planet-to-star surface brightness ratio $f_\mathrm{p}/f_\star$ - and sampled the posterior distributions using \emph{emcee}, as performed for the basic characterisation run of the system.

No significant occultation signals were identified. Based on the simulation results, we were able to set a 95\,\%~confidence upper limit on the planet-to-star surface brightness ratio of $(f_\mathrm{p}/f_\star)_\mathrm{max}$\,=\,$0.001$, which is compatible with the photometric precision. Given the planet-to-star area ratio $R_\mathrm{P}^{2}/R_{\star}^{2}=0.009849\pm0.000051$ (Table~\ref{Par_table}), this translates into an upper limit on the depth of the secondary eclipse of $\Delta I_{max}$=$(f_\mathrm{p}/f_\star)_\mathrm{max}$\,$R_\mathrm{P}^{2}/R_{\star}^{2}$$\approx$10~ppm.

The upper limit of the planet-to-star surface brightness ratio $(f_\mathrm{p}/f_\star)_\mathrm{max}$\,=\,$0.001$ allows us to constrain the upper limits of the geometric $A_\mathrm{g}$ and Bond $A_\mathrm{B}$ albedo of the planet. From the system parameters listed in Table~\ref{Par_table}, i.e., zero-eccentricity, effective stellar temperature \teff\,$=5520\pm60$~K, scaled semi-major axis $a_\mathrm{p}/R_{\star}=9.764\pm0.055$, and assuming $A_\mathrm{g} = 1.5\,A_\mathrm{B}$ and heat redistribution factor between 1/4 and 2/3, we found that $A_\mathrm{g}\le0.087\pm0.008$ and $A_\mathrm{B}\le0.058\pm0.006$. This makes Kepler-77b one of the gas-giant planets with the lowest albedo known to date, falling into range of albedos of \object{TrES-2b} \citep{Kipping2011} and \object{HD\,209458b} \citep{Rowe2008}.

\subsection{Search for additional planets}
\label{AdditionalPlanets}

We performed a photometric search for additional planets in the Kepler-77 system.

\textbf{Additional transits:} We searched the photometric data for additional transiting companions in the system. For this purpose, we used the techniques of \citet{Ofir2013a} on all the publicly available \kepler\ data. We cleaned the light curve by removing all in-transit data points of Kepler-77b and remaining transient effects, and used the optimal box least-squares (BLS) algorithm \citep{Ofir2013b} to search for additional transit signals. No additional transit signals were detected, in accordance with the trend that close-in giant planets do not tend to be found in transiting multi-planet systems \citep[e.g.,][]{Batalha2013}.

\textbf{Transit-timing variation:} Transits of a planets perturbed by additional objects are not strictly periodic. We carried out a search for dynamically induced deviations from a constant orbital period, i.e., we searched for variations in the transit centre times of Kepler-77b (transit-timing variations, TTVs). The search was carried out with an MCMC method similar to that used to characterise the planetary system (Sect~\ref{GlobalAnalysis}). We expanded the light curve model to include each time of central transit as a free parameter, yielding a model with a total of 308 parameters. The parameter priors were based on the posterior densities of the basic characterisation run. We approximated the posteriors of the characterisation run parameters with a normal distribution and used uninformative uniform priors on the transit centre times. The sampling was carried out with \emph{emcee} using 2000 parallel chains (walkers) and a notably longer burn-in phase.

\begin{figure}[t]
\centering
\resizebox{\hsize}{!}{\includegraphics[angle=0]{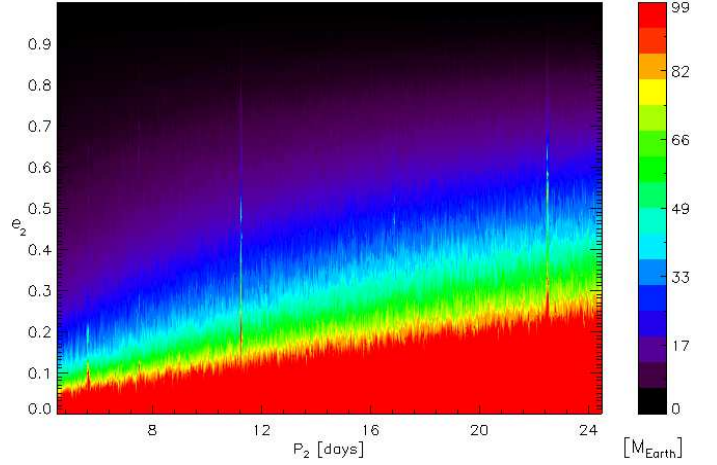}}
\caption{Maximum possible mass for a hypothetical perturber object in the Kepler-77 system as a function of its orbital period ($P_2$, in days) and eccentricity ($e_2$). Colours encode the upper limits of its mass (in Earth masses).}
\label{KOI127_TTVs_Analysis}
\end{figure}

The result of our TTVs search is shown in Fig.~\ref{KOI127_TTVs}, where the differences between the observed and modelled transit centre times - the so-called O$-$C diagram - are plotted against the transit number. We found that the transit centres do not deviate significantly from the linear ephemeris. The TTVs have a standard deviation of $\sim$0.5 minutes, with an absolute amplitude not exceeding $\sim$2~minutes. This value agrees with the result of \citet{Ford2011}, who, based upon the Q$_{1}$ and Q$_{2}$ \kepler\ data, found a maximum absolute variation of $\sim$1.8~minutes for Kepler-77b.

The lack of detectable TTVs in the Q$_{1}$--Q$_{13}$ \kepler\ light curve can be used to exclude certain types of companions. We used the approach described in \citet{Csizmadia2010} to give upper limits on the mass of a hypothetical perturbing object. Thirty thousand independent values of its orbital period and eccentricity were randomly chosen. Using the formalism of \citet{Agol2005}, at each given period and eccentricity, we calculated the perturber's maximum allowable mass that would cause TTV-amplitude shorter than 0.5~minutes (i.e., $1\,\sigma$). This formalism takes only coplanar orbits into account. Any mass-period-eccentricity combination that would cause a larger TTV-amplitude in the observational window considered in this paper ($\sim$3~years) can be excluded. The upper limits for the hypothetical perturber in a coplanar orbit are shown in Fig.~\ref{KOI127_TTVs_Analysis} as a function of its orbital period ($P_2$) and eccentricity ($e_2$). If there were a perturber on a short-period ($P_2\lessapprox24$\,days) and low-eccentricity ($e_2\lessapprox0.1$) orbit, it probably has a mass smaller than $\sim$100~M$_\mathrm{Earth}$. Low-mass ($M_2\le17$~M$_\mathrm{Earth}$) planets on highly eccentric ($e_2\ge0.3$) coplanar orbits cannot be ruled out with the currently available photometry. The possible existence of such planets should be studied in more details by stability investigations, which is beyond the scope of the present paper.

Planetary systems hosting hot-Jupiters are frequently considered as single-planet systems because no additional sub-stellar companions are detected, with only a handful of them showing TTVs \citep[e.g.,][]{Szabo2012}. Short time-scale TTVs are rarely present in systems harbouring close-in giant planets, and Kepler-77 agrees with this trend. A remarkable exception is \object{WASP-12b} \citep{Maciejewski2013}. The observed TTVs in some other systems hosting hot-Jupiters are often due to systematic observational effects \citep{Szabo2012}, stellar activity \citep{Barros2013}, and small-number statistics \citep{Fulton2011,Southworth2012}, rather than gravitational perturbations.

\subsection{Search for spot-crossing events: constraints on the spin-orbit obliquity}
\label{SpotsPlages}

Occultations of active photospheric regions by transiting planets can be used to constrain the spin-orbit obliquity of planetary systems \citep[see, e.g.,][]{Nutzman2011,Sanchis-Ojeda2011,Sanchis-Ojeda2012}, i.e., the angle between the stellar spin axis and the angular momentum vector of the orbit. The spin-orbit obliquity of planetary systems hosting close-in ($a\lessapprox0.15$~AU) gas giant planets is considered a keystone to investigate their formation, migration, and subsequent tidal interaction \citep[see, e.g.,][]{Winn2010, Morton2011a, Gandolfi2012}.

We searched for photometric anomalies in the transit light curves of Kepler-77 resulting from the passage of the planet in front of Sun-like spots. The search was carried out by removing the best-fitting transit signal from the 106 SC transit light curves and looking for systematic features in the residuals. The search was performed two-dimensionally by looking for features occurring during individual transits (planet crossing over a single feature) and over multiple transits (planet crossing over the same, possibly changing, feature over multiple transits).

\begin{figure}
\centering
\resizebox{8cm}{!}{\includegraphics[angle=0]{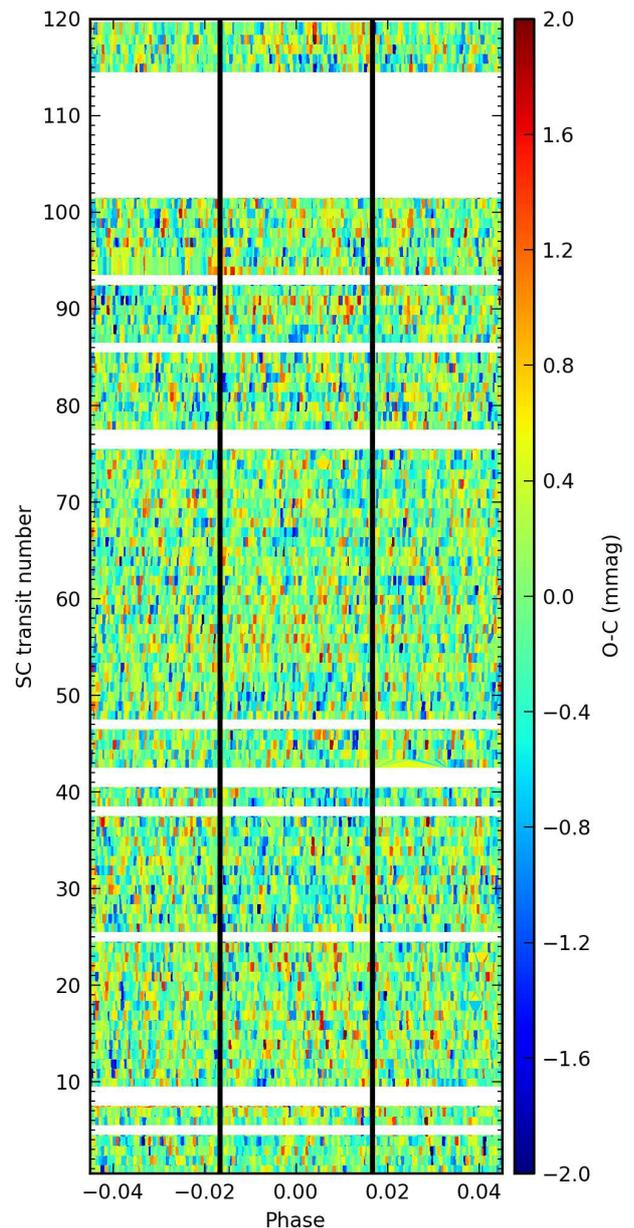}}
\caption{Short-cadence transit residuals map of Kepler-77, following the subtraction of the best-fitting transit model. Orbital phase increases from left to right, SC transit number from bottom to top. The two vertical thick lines mark the transit first and fourth planet's contact. The white bands, as well as the diagonal features in the residual map are caused by the lack of \kepler\ photometric data.}
\label{TransitResidualMap}
\end{figure}

No significant anomalies were identified above the noise level (1398\,ppm\,$\approx$\,1.5\,mmag) in the $\sim$1.2-year time window covered by the SC data, as shown in Fig.~\ref{TransitResidualMap}. The transits show marginal variations in the overall depth, which could be attributed to different levels of spot-coverage in the non-transited parts of the visible stellar disk, as suggested by the out-of-transit modulation visible in the light curve of Kepler-77 (Fig.~\ref{LongCadence-LightCurve}). 

\begin{figure*}[th]
  \centering
  \resizebox{15cm}{!}{\includegraphics[angle=0]{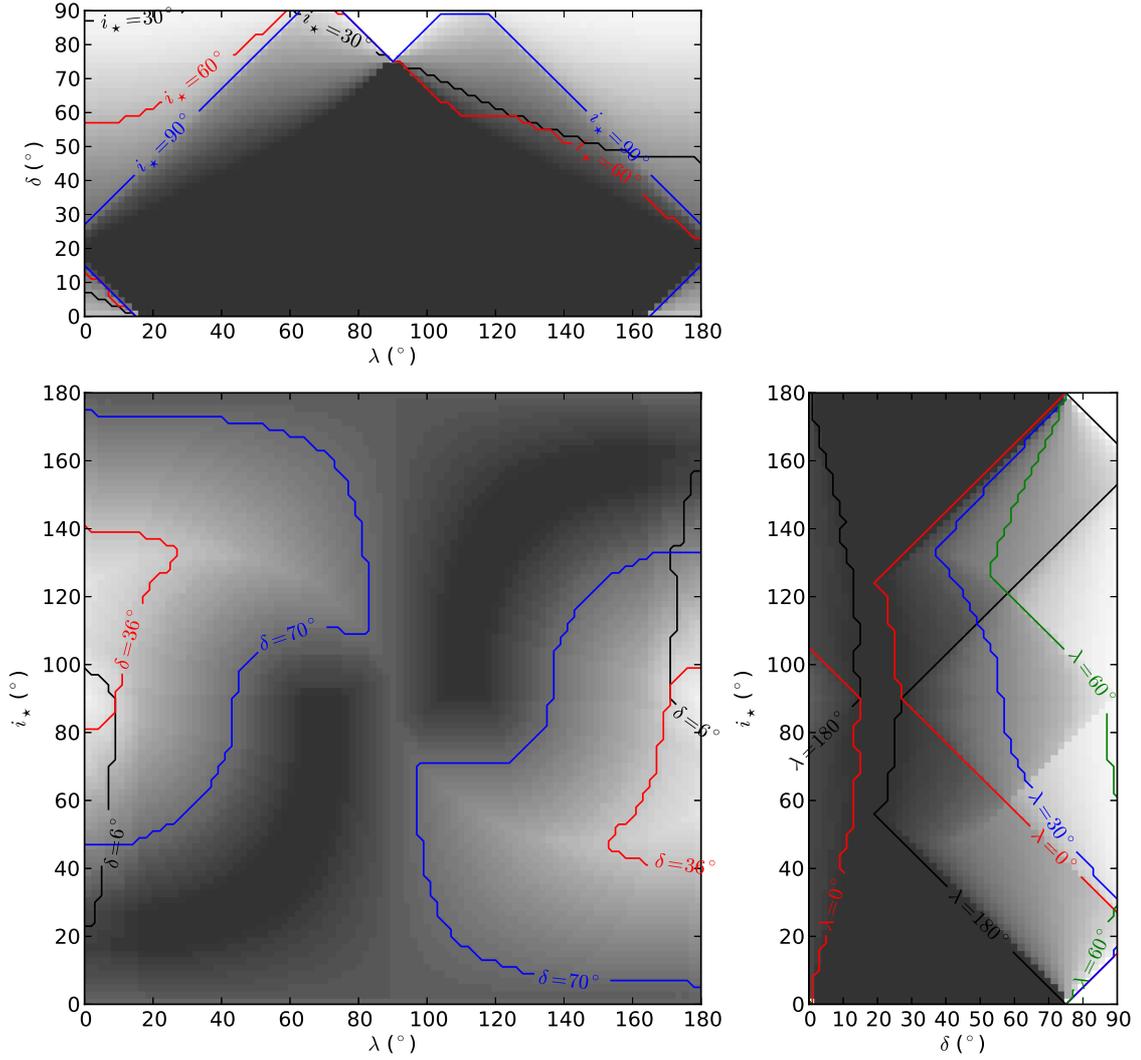}}
  \caption{Two-dimensional projections of the joint probability distribution for the stellar inclination $i_{\star}$, the characteristic latitude of active regions $\delta$, and the sky-projected spin-orbit angle $\lambda$ (see text for details).}
  \label{KOI127_grid}
\end{figure*}

The effective temperature of spots in mid G-type stars is estimated to be $\sim$1500\,K cooler than the unperturbed photosphere \citep[see, e.g.,][]{Strassmeier2009}. With a SC photometric precision of 1398~ppm, we can safely exclude occultations of starspots whose radius is larger than $\sim$0.04 stellar radii (assuming circular shape and neglecting limb-darkening, geometrical foreshortening, and penumbra/umbra effects). The $\sim$0.4\,\% out-of-transit stellar variability (Fig.~\ref{LongCadence-LightCurve}) indicates the presence of spots with a radius of $\sim$0.08 stellar radii. This suggests that the non-detection of significant anomalies is rather due to the non-intersection between the transit chord and the active latitudes of the star, and not to the lack of photometric precision.

In the Sun active regions tend to emerge predominantly at specific latitudes, which are symmetrically placed with respect to the solar equator. Over the course of the Sun's 11-year activity cycle, active latitudes gradually migrate from the mid-latitude (about $\pm35\degr$) towards the equator (about $\pm5\degr$), giving rise to the well-known butterfly diagram \citep[see, e.g.,][]{Li2001}. The spectral type, age, and photometric modulation of Kepler-77 are fairly similar to those of the Sun. It therefore seems reasonable to assume that photospheric active regions in Kepler-77 also fall primarily within two latitude bands that are symmetric with respect to the stellar equator. If we assume that spots on Kepler-77 were also confined to a given active latitude during the $\sim$1.2 years observations analysed here, then we can use the absence of spot-crossing events to place a joint constraint on the stellar inclination $i_{\star}$, the sky-projected spin-orbit obliquity $\lambda$, and the characteristic absolute latitude of the spots $\delta$.

To map out the region of $(i_{\star},\lambda,\delta)$-space that is excluded by the lack of detected spot crossings, we tracked a hypothetical point-like spot with latitude $\delta$ as the star rotates, and checked whether it overlapped with the transit chord at any point. If so, the particular combination of $i_{\star}$, $\lambda$, and $\delta$ was considered ruled out. We then repeated the process over a grid of values ranging from $0$ to $180\degr$ for $i_{\star}$ and $\lambda$, and from $0$ to $90\degr$ for $\delta$. Details of the calculation are given in Appendix~\ref{Appendix}. Note that considering this domain only does not imply any loss of generality; we are effectively assuming that if there are spots at $+\delta$, there are also spots at $-\delta$, and acknowledging  that we are only sensitive to $i_{\star}\,\mathrm{mod}\,180\degr$ and to $|\lambda|$ (i.e., $i_{\star}$  and $i_{\star}+180\degr$ are indistinguishable, as are $\lambda$ and $-\lambda$). Note also that we imposed no constraints on the stellar inclination $i_{\star}$, as the large relative error of the projected rotational velocity (\vsini\,$=1.5\pm1.0$~\kms) does not allow us to make a meaningful measurement of $i_{\star}$.

Figure~\ref{KOI127_grid} shows the two-dimensional projections of the three-dimensional grid; lighter regions mean that most combinations of the quantities along the $x$- and $y$-axis are permitted, and darker regions mean that most such combinations are excluded. The contours mark the edges of the permitted/excluded regions for specific values of the third angle. The predominantly grey area surrounded by blue contour in the upper left panel of Figure~\ref{KOI127_grid} shows the region of the $(\lambda,\delta)$-space that is excluded by the absence of spot-crossing event during the transits if the star is seen equator on. As the contours show, this region changes only slowly as the stellar inclination decreases, i.e., the constraints remain more or less unchanged if the star is seen close to, but not exactly equator-on. If active regions in Kepler-77 appeared within the same absolute latitude range observed in the Sun ($\sim$5--35$\degr$), for any value of the stellar inclination $i_{\star}$, the lack of starspots occultations in the transit light curve of Kepler-77 would imply $|\lambda| \lessapprox 26\degr$ or $154\degr\lessapprox\lambda\lessapprox206\degr$ (Figure~\ref{KOI127_grid}, middle left panel).

The SC \kepler\ data available at the time of writing cover $\sim$1.2 years. If Kepler-77 possesses an activity cycle similar to the Sun's, during which its active regions migrate in latitude, then starspots might drift into the range where they overlap with the transit chord over the next few years. Occultations of starspots by the planet would then become visible during its transits. This would allow us to study the magnetic cycle of the star and put stronger constraints on the spin-orbit obliquity of the system.

\begin{acknowledgements}
We are extremely grateful to the staff members at McDonald Observatory and Nordic Optical Telescope for their valuable and unique support during the observations. We also thank the anonymous referee for her/his careful review and suggestions. Davide Gandolfi wishes to thank John Kuehne, David Doss, Kate Isaak, Kevin Meyer, Ross Falcon, Michael Endl, Barbara Castanheira, Sydney Barnes, and John Southworth for the interesting and stimulating conversations. Hannu Parviainen has received support from the Rocky Planets Around Cool Stars (RoPACS) project during this research, a Marie Curie Initial Training Network funded by the European Commission's Seventh Framework Programme. He has also received funding from the V\"ais\"al\"a Foundation through the Finnish Academy of Science and Letters during this research. The IAC team acknowledges funding by grant AYA2010-20982 of the Spanish Ministry of Economy and Competitiveness (MINECO). Amy McQuillan's research has received funding from the European Research Council under the EU's Seventh Framework Programme (FP7/(2007-2013)/ ERC Grant Agreement No.~291352). Funding for the Stellar Astrophysics Centre is provided by The Danish National Research Foundation. The research is supported by the ASTERISK project (ASTERoseismic Investigations with SONG and \kepler) funded by the European Research Council (Grant agreement No.~267864). This research has made use of the Simbad database and the VizieR catalogue access tool, CDS, Strasbourg, France. The original description of the VizieR service was published in \aaps, 143, 23.

\end{acknowledgements}


\begin{table*}
\centering
\caption{Kepler-77 system parameters.}            
\begin{tabular}{l r}
\hline
\hline
\noalign{\smallskip}
\multicolumn{2}{l}{\emph{Model parameters}} \\
\noalign{\smallskip}
\hline
\noalign{\smallskip}
Planet orbital period $P_\mathrm{orb}$ [days]                       &  $3.57878087\pm0.00000023$  \\
\noalign{\smallskip}
Planetary transit epoch $T_\mathrm{c}$ [HJD-2\,450\,000]            &  $5095.865727\pm0.000029$   \\
\noalign{\smallskip}
Planetary transit duration $T_{14}$ [days]                          &  $0.12224\pm0.00014$        \\
\noalign{\smallskip}
Planet-to-star area ratio $R_\mathrm{p}^{2}/R_{\star}^{2}$          &  $0.009849\pm0.000051$      \\
\noalign{\smallskip}
Impact parameter $b$                                                &  $0.341\pm0.017$            \\
\noalign{\smallskip}
Linear limb-darkening coefficient $u_1$                             &  $0.505\pm0.015$            \\
\noalign{\smallskip}
Quadratic limb-darkening coefficient $u_2$                          &  $0.139\pm0.032$            \\
\noalign{\smallskip}
\kepler\ LC data scatter $\sigma_\mathrm{LC}$ [ppm]                 &  $378.1\pm3.8$              \\
\noalign{\smallskip}
\kepler\ SC cadence data $\sigma_\mathrm{SC}$ [ppm]                 &  $1397.5\pm3.6$             \\
\noalign{\smallskip}
Radial velocity semi-amplitude $K$ [\ms]                            & $ 59.2\pm4.3 $              \\
\noalign{\smallskip}
Systemic velocity (Sandiford) $V_{\gamma\,\mathrm{Sand}}$ [\kms]    & $-24.7843\pm0.0120$         \\
\noalign{\smallskip}
Systemic velocity (FIES) $V_{\gamma\,\mathrm{FIES}}$ [\kms]         & $-24.7567\pm0.0035$         \\
\noalign{\bigskip}
\multicolumn{2}{l}{\emph{Derived parameters}} \\
\noalign{\smallskip}
\hline
\noalign{\smallskip}
Planet-to-star radius ratio $R_\mathrm{p}/R_{\star}$                   & $0.09924\pm0.00026$  \\
\noalign{\smallskip}
Scaled semi-major axis of the planetary orbit $a_\mathrm{p}/R_{\star}$ &   $9.764\pm0.055$    \\
\noalign{\smallskip}
Semi-major axis of the planetary orbit $a_\mathrm{p}$ [AU]             & $0.04501\pm0.00063$  \\
\noalign{\smallskip}
Orbital inclination angle $i_\mathrm{p}$ [degree]                      &   $88.00\pm0.11$     \\
\noalign{\smallskip}
Orbital eccentricity $e$                                               &       0 (fixed)      \\
\noalign{\smallskip}
Ingress -- egress duration $T_{12}=T_{34}$                             & $0.02478\pm0.00036$  \\
\noalign{\smallskip}
Bulk stellar density $\rho_{\star}$ [\gcm3]                            &   $1.365\pm0.023$    \\
\noalign{\bigskip}
\multicolumn{2}{l}{\emph{Stellar fundamental parameters}} \\
\noalign{\smallskip}
\hline
\noalign{\smallskip}
Effective temperature $T_\mathrm{eff}$ [K]                             &   $5520\pm60$        \\
\noalign{\smallskip}
Surface gravity\tablefootmark{a} log\,$g$ [dex]                        &   $4.40\pm0.10$      \\
\noalign{\smallskip}
Surface gravity\tablefootmark{b} log\,$g$ [dex]                        &   $4.42\pm0.01$      \\
\noalign{\smallskip}
Metallicity $[\mathrm{M/H}]$ [dex]                                     &   $0.20\pm0.05$      \\
\noalign{\smallskip}
Microturbulent velocity\tablefootmark{d} $v_ {\mathrm{micro}}$ [\kms]  &    $0.9\pm0.1$       \\
\noalign{\smallskip}
Macroturbulent velocity\tablefootmark{d} $v_ {\mathrm{macro}}$ [\kms]  &    $1.8\pm0.3$       \\
\noalign{\smallskip}
Projected stellar rotational velocity \vsini\ [\kms]                   &    $1.5\pm1.0$       \\
\noalign{\smallskip}
Spectral type\tablefootmark{c}                                         &       G5\,V          \\
\noalign{\smallskip}
Star mass $M_{\star}$ [\Msun]                                          &   $0.95\pm0.04$      \\
\noalign{\smallskip}
Star radius $R_{\star}$ [\Rsun]                                        &   $0.99\pm0.02$      \\
\noalign{\smallskip}
Star age $t$ [Gyr]                                                     &    $7.5\pm2.0$       \\ 
\noalign{\smallskip}
Star rotation period $P_\mathrm{rot}$ [days]                           &     $36\pm6$         \\
\noalign{\smallskip}
Interstellar extinction $A_\mathrm{V}$ [mag]                           &   $0.08\pm0.04$      \\
\noalign{\smallskip}
Distance of the system $d$ [pc]                                        &    $570\pm70$        \\
\noalign{\bigskip}
\multicolumn{2}{l}{\emph{Planetary fundamental parameters}} \\
\noalign{\smallskip}
\hline
\noalign{\smallskip}
Planet mass\tablefootmark{e} $M_\mathrm{p}$ [\Mjup]                    & $0.430\pm0.032  $    \\
\noalign{\smallskip}
Planet radius\tablefootmark{e} $R_\mathrm{p}$ [\Rjup]                  & $0.960\pm0.016  $    \\
\noalign{\smallskip}
Planet density $\rho_\mathrm{p}$ [\gcm3]                               & $0.603\pm0.055  $    \\
\noalign{\smallskip}
Equilibrium temperature\tablefootmark{f} $T_\mathrm{eq}$ [K]           & $1440^{+100}_{-120}$ \\
\noalign{\smallskip}
Geometric albedo  $A_\mathrm{g}$                                       & $\le0.087\pm0.008$   \\ 
\noalign{\smallskip}
Bond albedo $A_\mathrm{B}$                                             & $\le0.058\pm0.006$   \\ 
\noalign{\smallskip}
\hline       
\end{tabular}
\tablefoot{~\\
  \tablefoottext{a}{Obtained from the spectroscopic analysis.}\\
  \tablefoottext{b}{Obtained from $T_\mathrm{eff}$, $[\mathrm{M/H}]$, and $\rho_{\star}$, along with the Pisa Stellar Evolution Data Base for low-mass stars.}\\
  \tablefoottext{c}{With an accuracy of $\pm\,1$ sub-class.}\\
  \tablefoottext{d}{Using the calibration equations of \citet{Bruntt2010}.}\\
  \tablefoottext{e}{Radius and mass of Jupiter taken as 7.1492$\times10^9$~cm and 1.89896$\times$10$^{30}$~g, respectively.}\\
  \tablefoottext{f}{Assuming $A_\mathrm{B}\le0.058\pm0.006$ and heat redistribution factor between 1/4 and 2/3.}
}
\label{Par_table}  
\end{table*}

\begin{appendix}
\section{Tracking a rotating starspot with respect to the transit chord}
\label{Appendix}

\begin{figure}[t]
  \centering
  \resizebox{\hsize}{!}{\includegraphics[angle=0]{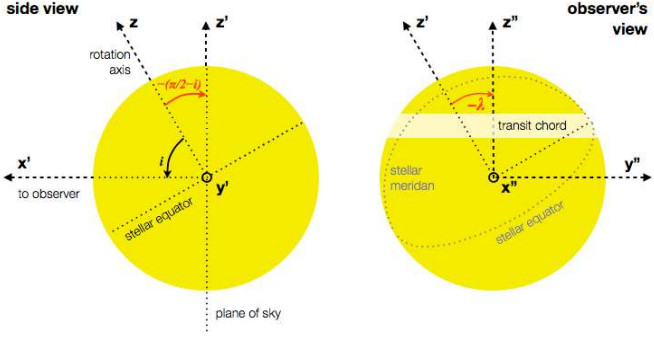}}
  \caption{System geometry seen from the side (left) and from the observer's standpoint (right), illustrating the two rotations needed to convert from frame $S$ to frame $S''$ (see text for details).}
  \label{fig:frames}
\end{figure}

The geometry of the system is shown in Figure~\ref{fig:frames}. We started by considering a right-handed reference frame $S$, whose origin coincides with the centre of the star. In this frame, the stellar rotation vector defines the $+z$-direction, while the $y$-axis lies in the plane of the sky. If we set the stellar radius to unity, the coordinates of a starspot with latitude $\delta$ in this reference frame are simply
\begin{displaymath}
  \left[
    \begin{array}{l}
      x \\
      y \\
      z 
    \end{array} 
  \right]
  = 
  \left[
    \begin{array}{l}
      \cos \delta \cos \phi  \\
      \cos \delta \sin \phi  \\
      \sin \delta
    \end{array} 
  \right]\,,
\end{displaymath}
where $\phi = 2 \pi t / P_\mathrm{rot} + \phi_0$, $P_\mathrm{rot}$ is the star's rotational period, and $\phi_0$ is the longitude of the spot at some arbitrary time $t=0$. 

We defined a second right-handed reference frame $S'$ that shares the same origin and $y$-axis as frame $S$ (i.e., $y' \equiv y$), but whose $x'$-axis is pointing towards the observer. The stellar inclination $i_{\star}$ is the angle between the stellar rotation axis and the line-of-sight, i.e., the angle in the $x'z'$-plane between the $z$-axis of frame $S$ and the $x'$-axis of frame $S'$. Coordinates in frame $S$ are transformed to frame $S'$ by performing a rotation by $-(\pi/2-i_{\star})$ about the $y$-axis (or, equivalently, around the $y'$-axis; see Figure~\ref{fig:frames}, left panel). This process is described in more detail in Appendix~A of \citet{Aigrain2012}. 

We then considered a third right-handed reference frame $S''$ that shares the same origin as $S$ and $S'$, and the same $x$-axis as frame $S'$ (i.e., $x'' \equiv x'$), but has its $y''$-axis parallel to the transit chord. The sky-projected spin-orbit angle $\lambda$ is defined as the angle in the plane of the sky between the projections of the orbital angular momentum and of the stellar spin axis, i.e., the angle in the $y''z''$-plane between the $z''$-axis of frame $S''$ and the $z'$-axis of frame $S'$. Coordinates in frame $S'$ are transformed to frame $S''$  by performing a second rotation by $-\lambda$ about the $x'$-axis (or, equivalently, around the $x''$-axis; see Figure~\ref{fig:frames}, right panel). The coordinates of the spot in frame $S''$ are thus 
\begin{displaymath}
  \centering
   \begin{array}{rcl}
    \left[
      \begin{array}{l}
        x'' \\
        y'' \\
        z'' 
      \end{array} 
    \right]
    & = & 
    \left[
      \begin{array}{l}
        \cos \delta \cos \phi \sin i_{\star} + \sin \delta \cos i_{\star} \\
        \cos \lambda \cos \delta \sin \phi + \sin \lambda \cos
        \delta \cos \phi \cos i_{\star} - \sin \lambda \sin \delta \sin i_{\star}
        \\
        \sin \lambda \cos \delta \sin \phi - \cos \lambda \cos
        \delta \cos \phi \cos i_{\star} + \cos \lambda \sin \delta \sin i_{\star}.
      \end{array} 
    \right].
  \end{array}
\end{displaymath}
If we observe the system for an infinite amount of time, a given spot that remains static relative to the rotating surface of the star will eventually be crossed by the planet during a transit, if, and only if,
\begin{displaymath}
x''>0~{\rm and}~b - R_\mathrm{p} / R_\star < z'' < b + R_\mathrm{p} / R_\star.
\end{displaymath}
The absence of spot crossings therefore excludes a specific volume in $(i_{\star},\lambda,\delta)$-space that we explored by following the variations of $x''$ and $z''$ as $\phi$ varies. In the present study, this was done by stepping through a grid of values, ranging from $0$ to $180\degr$ for $i_{\star}$ and $\lambda$, and from $0$ to $90\degr$ for $\delta$, with $2\degr$ steps. At each point of the grid, $x''$ and $z''$ were evaluated over a grid of $\phi$-values, ranging from $0$ to $360\degr$ with $2\degr$ steps. If the conditions above (evaluated using the values for $b$ and $R_\mathrm{p}/R_{\star}$ given in Table~\ref{Par_table}, i.e., 0.341 and 0.09924, respectively) were met for any value of $\phi$, the corresponding cell in the $(i_{\star},\lambda,\delta)$ grid was considered excluded.

\end{appendix}

\end{document}